# Automatic extraction of coronary arteries using deep learning in invasive coronary angiograms

Yinghui Meng[1], Zhenglong Du[1], Chen Zhao[2], Minghao Dong[1], Drew Pienta[3], Zhihui Xu[4], Weihua Zhou[2,5]

[1]School of Computer and Communication Engineering, Zhengzhou University of Light Industry, Zhengzhou, Henan, China

[2]Department of Applied Computing, Michigan Technological University, Houghton, MI, 49931, USA

[3]DepartmentMechanical Engineering-Engineering Mechanics, Michigan Technological University, Houghton, MI, 49931, USA

[4]Department of Cardiology, The First Affiliated Hospital of Nanjing Medical University, Nanjing, Jiangsu, 210000, China

[5]Center for Biocomputing and Digital Health, Institute of Computing and Cybersystems, and Health Research Institute, Michigan Technological University, Houghton, MI, 49931, USA

**Abstract**

**Background.** Accurate extraction of coronary arteries from invasive coronary angiography (ICA) is important in clinical decision-making for the diagnosis and risk stratification of coronary artery disease (CAD). In this study, we develop a method using deep learning to automatically extract the coronary artery lumen.

**Methods.** A deep learning model U-Net 3+, which incorporates the full-scale skip connections and deep supervisions, was proposed for automatic extraction of coronary arteries from ICAs. Transfer learning and a hybrid loss function were employed in this novel coronary artery extraction framework.

**Results.** A data set containing 616 ICAs obtained from 210 patients was used. In the technical evaluation, the U-Net 3+ achieved a Dice score of 0.8942 and a sensitivity of 0.8735, which is higher than U-Net ++ (Dice score: 0.8814, the sensitivity of 0.8331) and U-net (Dice score: 0.8799, the sensitivity of 0.8305).

**Conclusion.** Our study demonstrates that the U-Net 3+ is superior to other segmentation frameworks for the automatic extraction of the coronary arteries from ICAs. This result suggests great promise for clinical use.

# 1. Introduction

Coronary artery disease (CAD) is a disease of the coronary vessels, including angina pectoris and acute myocardial infarction, which is the leading cause of mortality in developing countries [1][2][3]. As atherosclerotic plaque builds up in the epicardial arteries, it may lead to coronary artery stenosis. Coronary artery stenosis causes to a starvation of relative myocardial oxygen supply and commonly results in ischemia [4]. Severe myocardial ischemia, which can lead to serious symptoms such as angina or even myocardial infarction [5].

Invasive coronary angiography (ICA), which provides assessments of artery stenosis and plaque characteristics, is an essential tool for CAD evaluation and treatment [6]. Automated extraction of coronary arteries from ICAs is the first step before stenosis detection, which plays a key role in the clinical practice of interventional cardiology. However, the accurate extraction of the coronary arteries from ICAs is a challenging task for the following reasons: 1) complex noise caused by the non-uniform illumination; 2) poor signal to noise ratio; 3) uneven intensity; 4) semantic information confusion. Automatic extraction techniques of blood vessels can be broadly classified into two main categories: traditional image processing and deep learning-based methods. The former includes Gaussian or Gabor filter-based, model-based, and line tracking-based methods [7],[8],[9], which focus primarily on the vascular extraction of low noise and high-contrast. However, these methods cannot achieve satisfactory results when dealing with ICAs with extremely complex noise. The latter mainly contains convolution neural network-based methods (CNNs), which have shown excellent feature extracting performance. Yang designed a method that applied correspondence matching and CNN for automatic coronary artery extraction, and a Dice score of 0.8007 was reported [10]. Nasr-Esfahani reported that using patches of pixels to evaluate the ICAs and a CNN was designed to determine the arteries and background [11]. To further improve the segmentation accuracy, E. Nasr et al. used two CNNs to a combination of primary and secondary features and achieved a Dice score of 0.8151 [12]. Image segmentation based on CNN methods, such U-Net and U-Net++, has also made outstanding achievements [13]. Zhao et al. proposed FP-U-Net++ by introducing the feature pyramid techniques into U-Net++ and reached a dice score of 0.8899 [14],[15]. However, these methods lack the ability to explore enough information from the full scale. The U-NET 3+ [16] overcomes this drawback by utilizing full-scale skip connections and deep supervision.

The objective of this study was to demonstrate the use of the U-NET 3+ algorithm to automatically extract the coronary arteries from ICAs.

## 2. Materials and methods

### 2.1. Patient data

This study was approved by the ethics committee of The First Affiliated Hospital of Nanjing Medical University. 210 patients (100 males and 110 females) who received ICA (616 images) were retrospectively analyzed. The acquired ICA images were 512 × 512 pixels per scan, with pixel spacing ranging from 0.200-0.390 mm.

### 2.2 Artery extraction

The deep learning model U-Net 3+ is used to extract the coronary artery from ICAs (Fig.1 b). ICA images (Fig. 1a) are fed into the U-Net 3+ network (Fig.1 b), and the output are arterial contours (Fig. 1c).

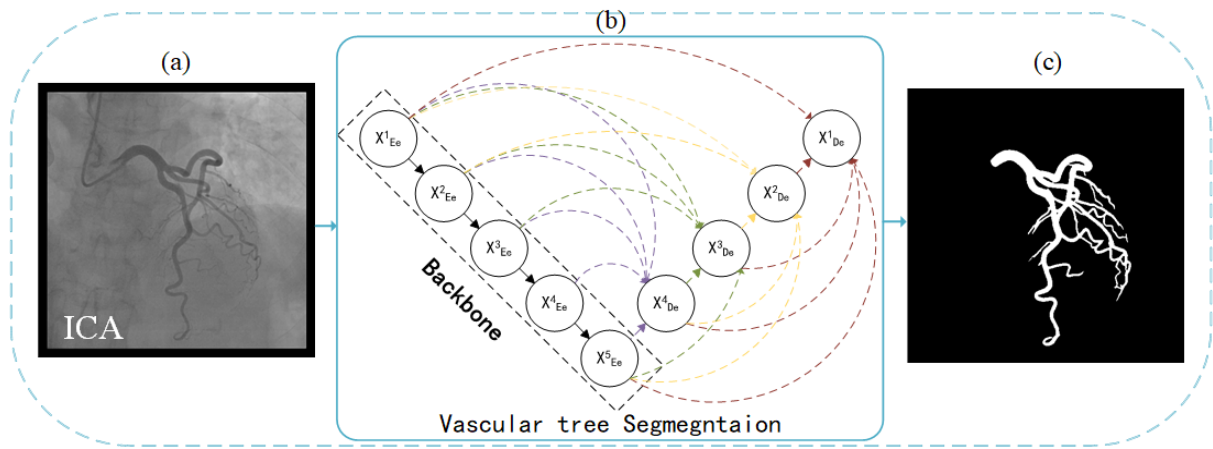

**Fig. 1** Workflow of extraction coronary artery. (a) ICA image. (b) U-Net 3+ network. (c) Predicted arterial contour.

### 2.2.1 Deep neural network architecture

The simplified overview of U-Net is illustrated in Figure 2(a). The network architecture consists of encoder and decoder paths. The encoder path is on the left, and the decoder is on the right. The low-level extracted feature maps from the encoder path are directly concatenated into the corresponding decoder path. The nodes $X^{i,j}$ represent the result of the convolutional layers, where $i$ is the index of the encoder and $j$ is the index of the decoder. The decoder part of the feature maps represented by $X^{i,j} (i = 4-j)$ is computed in Eq. (1).

$$X^{i,j} = \begin{cases} F(D(X^{i-1,j})) & j = 0 \\ F([X^{i,0}, U(X^{i+1,j-1})]) & j > 0 \end{cases} \quad (1)$$

Where $F(\cdot)$ realizes the convolution operation followed by a batch normalization and a ReLU activation function [17] and a ReLU activation function. The [·] denotes the concatenation operation. $D(\cdot)$ and $U(\cdot)$ represents the down-sampling operation and up-sampling operation, respectively.

U-Net++ is a modified solution of U-Net. U-Net++ upgrades the U-Net techniques by introducing the nested and dense skip connections. As shown in Figure 2(b), what separates U-Net++ from the original U-Net is that the former has one encoder and multiple decoders and undergo a dense convolution layers. In U-Net++, the decoder part of the feature maps represented by $X^{i,j}$ are computed in Eq. (2).

$$X^{i,j} = \begin{cases} F(D(X^{i-1,j})) & j = 0 \\ F([[X^{i,k}]_{k=0}^{j-1}, U(X^{i+1,j-1})]) & j > 0 \end{cases} \quad (2)$$

The decoding is not done directly from the encoding, and the number of layers in this convolution depends on the down-sampling layer corresponding to the coding direction [15]. The number of convolutional layers k is computed in Eq. (3).

$$k = 3 - i (i = 0, 1, 2, 3) \quad (3)$$

The simplified overview of U-Net 3+ is illustrated in Figure 2(c). U-Net 3+ has full-scale skip connections and deep supervisions. The encoder part is implemented by an advanced CNN classifier i.e., InceptionResNetV2 [18]. Different from U-net and U-net++, each decoder layer in U-Net 3+

incorporates both encoder and decoder feature maps from the same scale. The decoder part of the feature maps represented by $X_{De}$ is computed in Eq. (4).

$$X_{De}^i = \begin{cases} X_{En}^i & i = N \\ F\left(\left[\underbrace{C(D(X_{En}^k))_{k=1}^{i-1}, C(X_{En}^i)}_{scalest(i+1)^{th}}, \underbrace{C(U(X_{De}^k))_{k=i+1}^{N}}_{scalest1^{th} \sim i^{th}}\right]\right), & i = 1,...N-1 \end{cases} \quad (4)$$

Where $C(\cdot)$ represents the convolution operation.

The full-scale deep supervision in U-Net 3+ is used to supervise on each scale. The side outputs of the decoder i.e., $X_{Ee}^5, X_{De}^4, X_{De}^3$ and $X_{De}^2$ are fed into a plain 3 × 3 convolution layer and then up-sampled with a factor of 16, 8, 4 and 2, respectively. Subsequently, the up-sampled feature maps from $X_{Ee}^5, X_{De}^4, X_{De}^3$ and $X_{De}^2$ have the same spatial size as that from $X_{De}^1$. Moreover, there are converted into a probability map by sigmoid function. The network weights are optimized by comparing the differences between the ground truth of all the side outputs and the arterial annotations. Finally, the OTSU algorithm [19] is employed to convert the probability map into binary segmentation results where a pixel value of 1 is represented as coronary artery, and 0 is described as background.

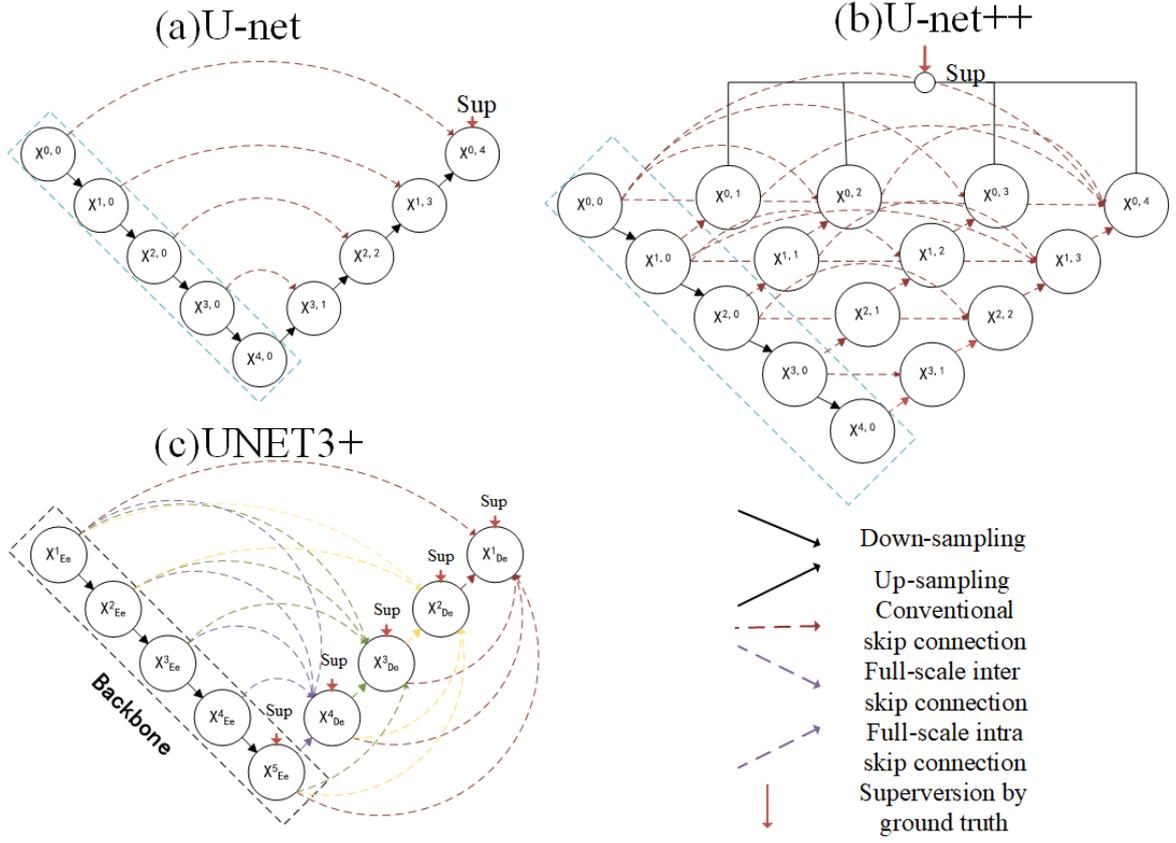

**Fig. 2**. Comparison of arterial extraction models of the (a) U-Net; (b) U-Net++; (c) U-Net 3+. In (c), the dashed line indicates the full-scale Skip Connections for U-Net 3+.

**2.2.2 Loss function**

In practice, training the model is to minimize the loss function in the process and thus obtain a higher Dice score (DSC). To train the U-Net3+ model smoothly, a hybrid loss function is employed to control backpropagation. It contains Binary Cross Entropy loss (BCE_loss) [20], dice loss, and an L2 regularization term. Cross entropy is one of the most common evaluation methods to determine the closeness between an output vector and the expectation vector. It describes the distance between two probability distributions, which is widely used in classification tasks [21]. Coronary artery extraction is a single-label binary classification task whose purpose is to divide arterial targets from ICAs. The BCE_loss is defined in Eq. (5).

$$Bce\_loss = y\log(\bar{y}) + (1-y)\log((1-\bar{y})) \qquad (4)$$

Where $y$ denotes the ground truth, $\bar{y}$ denotes the predicted binary arterial tree. The arterial contours are only a tiny fraction of the entire pixel in the ICAs. An ensemble similarity meter function, Dice, is introduced to alleviate the data sample discrepancy problem. It is usually used to calculate the similarity of two samples, and assign greater training weights to smaller samples in the dataset. The DSC is computed in Eq. (6).

$$DSC = \frac{2|\bar{y} \cap y|}{|\bar{y}| + |y|} \quad (5)$$

Where $|\bar{y} \cap y|$ denotes the intersection of sets $\bar{y}$ and $y$. $|\bar{y}|$ and $|y|$ denotes the number of their elements. The larger the DSC, the smaller the loss. A $DSC$ of 1 demonstrates that there is no difference between the ground truth and the predicted value, while a $DSC$ of 0 indicates complete dissimilarity.

A simple sample is one that can be quickly learned and accurately judged by the model. A complex sample refers to one with insignificant features that can be easily misjudged. The uneven distribution of simple and complex samples affects the training of deep CNNs. In the task of coronary artery extraction, there are many complex samples in the foreground class that are difficult to judge correctly, such as thinner blood vessel branches. To deal with the problem of the imbalanced distribution of difficulty in samples, it is necessary to make the model focus more attention on the learning of complex samples, which can be achieved by reducing the contribution of simple samples to the total loss and increasing the penalty weight of complex samples. Therefore, we utilize the overall loss defined as Eq. (7) as the loss function of the backbone network.

$$L(\theta)_{loss} = (1 - y\log(\bar{y}) + (1 - y)\log((1 - \bar{y}))) + \log(1 - \frac{2|\bar{y} \cap y|}{|\bar{y}| + |y|}) + \|W\|_2 \quad (6)$$

Due to the particularity caused by the low-contrast and high-noise of the ICAs of coronary artery extraction task, more attention should be paid to multi-scale information. The loss function should be used on each branch of deep supervision to control the training of the network. The deep supervised branching loss function is defined as Eq. (8).

$$L(\mathrm{s})_{loss} = (1-y\log(\overline{y})+(1-y)\log((1-\overline{y}))) + \log(1-\frac{2|\overline{y}\cap y|}{|\overline{y}|+|y|}) + \|W\|_2 \qquad (7)$$

The hybrid loss function is defined in Eq. (9). as the objective loss function.

$$L_{loss} = \frac{L(0)_{loss}+L(s)_{loss}}{s} \qquad (8)$$

The s is the number of layers of all lateral outputs of the depth neural network.

### 2.2.3 Optimization strategy

The U-Net 3+ used in this study was implemented in Python with a **PyTorch** backend. The CNN we utilized here is a supervised learning model and requires both data and ground truth labels. To further improve the accuracy of coronary extraction, transfer learning was applied. The advantages of transfer learning include better performance and reduced training time [22],[23]. By transfer learning, we could enable the CNN to retain the weights learned in the previous task and then apply them in the new task. Therefore, we introduced pre-trained weights from ImageNet into our network during the training process. This experiment demonstrated that the network performance was better when the encoder structure of U-Net 3+ is similar to that of InceptionResNetV2. During the training of the U-Net 3+ model, the initial learning rate was set to 0.0001 and a decay was set to 0.0001. The RMSProp optimizer [24] was employed to update the weights within the CNN.

The manually labeled 616 ICAs were still inadequate to train a 2D CNN. To solve this problem and enhance the stability of the CNN model, data augmentation techniques were applied. The techniques include flipping, rotating, scaling, cropping, shifting, Gaussian noise transformation, and Gaussian fuzzy transformation, respectively. Those were applied to make minor changes to the existing data sets randomly. The image obtained after data augmentation will be recognized as two different images by the deep learning model during the network training process, thus achieving the purpose of expanding the data set. When images were fed into the model for training, a random sampling strategy was applied, and each type of data enhancement technique was executed randomly. Also, if an image performs data enhancement, the label performs the same data enhancement technique accordingly.

## 2.3 Metrics to evaluate the accuracy of artery extraction

Five metrics were used to evaluate the accuracy of coronary artery extraction. There are DSC, sensitivity (SN), specificity (SP), Hausdorff distance (HD) and average surface distance (ASD).

SN denotes the ratio of pixels predicted to be positive samples to all true positive sample pixels, and SP denotes the ratio of pixels predicted as negative samples to all true negative sample pixels. Hausdorff distance represents the degree of similarity between two sets of points. Assume there are two sets of sets $C = \{a1......an\}$ and $D = \{b1......bn\}$. the Hausdorff distance between these two sets of points is defined as Eq. (10).

$$H(C,D) = \max(h(C,D), h(D,C)) \tag{9}$$

The definitions of $h(C,D)$ and $h(D,C)$ are given in Eq. (11) and Eq. (12).

$$h(C,D) = \max_{a \in C} \left\{ \min_{b \in D} \|a - b\| \right\} \tag{10}$$

$$h(C,D) = \max_{b \in D} \left\{ \min_{a \in A} \|b - a\| \right\} \tag{11}$$

The surface distance is a function describing the average variance in the surface between the prediction results and ground truth. Let P(A) indicates the set of surface pixels of A. The shortest distance of an arbitrary pixel $v$ to P(A) is defined as Eq. (13).

$$d(v, P(A)) = \min_{p_A \in P(A)} \|v - p_A\| \tag{12}$$

Where $\|.\|$ denote the Euclidean distance. The surface distance is defined as Eq. (14).

$$ASD = \frac{\sum_{a \in P(A)} \min_{b \in P(B)} \|a - b\| + \sum_{b \in P(B)} \min_{a \in P(A)} \|b - a\|}{|P(A)| + |P(B)|} \tag{13}$$

## 3. Results and analysis

### 3.1 Dataset

A total of 616 ICA images were used in our work, 405 of which were left coronary artery (LCA) images, and 211 were right coronary artery (RCA) images. As the basis sample used for deep learning network training, these images were manually annotated by well-trained operators. Table 1 contains specific information on the ICA images used in this paper.

We divided the dataset into the training set, validation set, and test set according to what ratios of 70%, 10%, and 20%. 210 patients had 616 ICA images, of which 437 images were used for the training set, 49 images were used for the validation set, and 130 images were used for the test set.

**Table1**

Indicators of the number of different angles of LCA and RCA. Note that LCA and RCA represents the left coronary artery and right coronary artery. LAO and RAO represent the left anterior oblique and right anterior oblique.

|     | LAO | RAO | Total |
| --- | --- | --- | --- |
| LCA | 165 | 240 | 405 |
| RCA | 163 | 48  | 211 |

### 3.2 Accuracy of artery extraction

DSC was used to evaluate the vessel extraction ability of the deep learning models applied in our paper on an external test set and used 10-fold cross-validation for the final evaluation.

Two different methods are compared with the tested U-Net 3+ including U-Net and U-Net++. However, the decoder part in all the three networks adopted the same structure as in InceptionRes-NetV2. U-Net 3+ without Deep Supervision (U-Net3+ DS-) and U-Net3+ without Deep Supervision and transfer learning (U-Net3+ DS- and TL-) of U-Net3+ were individually used as a comparison. In addition, four deep learning models, DeepLabV3 [25], DeepLabV3+ [26], PSPNet [27] and CGNet [28] were employed for comparison. Transfer learning and data augmentation were applied to all deep learning models.

As shown in Figure 3, examples of extracting arteries from ICAs by different deep learning models are delineated in red. We can observe that the U-Net3+ algorithm leads to the best arteries extraction with the closest visual appearance to the reference ground truth among all the methods.

In Table 2, the HD, DSC, Dice, SN, SP, and ASD indexes are given in terms of average scores for the 10-fold cross-validation results of the test dataset. The quantitatively compared results of the model performance of U-Net, U-Net++, U-Net3+ DS- and TL-, U-Net3+ DS-, DeepLabV3, DeepLabV3+, PSPNet, CGNet and U-Net 3+ under different metrics were presented. The average DSC for U-Net 3+, U-Net++, and U-Net are 0.8942, 0.8814, and 0.8799, respectively. The results show in Table 2 that the U-Net 3+ algorithm obtains the lowest HD and ASD and achieves the highest scores of DSC. In addition, four existing studies for coronary artery segmentation are presented in Table 3.

As shown in Figure 4, 10-fold cross-validation was employed to ensure the stability of the experimental results. The DSC, SP and SN of each experiment was almost on the same level (DSC: min: 0.8913; max: 0.8980), so it could be concluded that our experimental results have good reliability and stability.

**Table 2**

Quantitative results associated with different methods for the images in the testing dataset. Note that HD and ASD represents the Hausdorff distance and average surface distance. DSC, SN and SP represents Dice similarity coefficient, sensitivity and specificity, respectively.

| Mertic | HD(mm) | DSC | SN | SP | ASD(mm) |
| --- | --- | --- | --- | --- | --- |
| CGNet | 7.6218 | 0.5395 | 0.4728 | 0.9898 | 6.7381 |
| PSPNet | 6.7776 | 0.8257 | 0.7908 | 0.9924 | 0.5668 |
| DeepLabV3 | 6.8468 | 0.8340 | 0.8178 | 0.9911 | 0.4950 |
| DeepLabV3+ | 6.5714 | 0.8462 | 0.8123 | 0.9937 | 0.7651 |
| U-Net | 6.1040 | 0.8799 | 0.8305 | 0.9967 | 0.4403 |
| U-Net++ | 6.1048 | 0.8814 | 0.8331 | 0.9966 | 0.4588 |
| U-Net3+w/o DS and TL | 6.2257 | 0.8607 | 0.8160 | 0.9958 | 0.8840 |
| U-Net3+w/o DS | 6.1024 | 0.8862 | 0.8626 | 0.9955 | 0.5080 |
| U-Net3+ | **6.0794** | **0.8942** | **0.8735** | 0.9954 | **0.3247** |

**Table 3**

Comparison among the state-of-the-art methods. Note that DSC, SN and SP represents Dice similarity coefficient, sensitivity and specificity, respectively.

| Mertic | Data（images） | DSC | SN | SP |
| --- | --- | --- | --- | --- |
| CNN[11] | 44 | 0.7562 | 0.7935 | 0.9890 |
| CNN[12] | 44 | 0.8151 | 0.8676 | 0.9859 |
| Multiple CNNs[29] | 120 |  | 0.7774 | 0.9934 |
| FP-U-Net++[14] | 314 | 0.8899 | 0.8595 | 0.9960 |
| U-Net 3+ | 616 | **0.8942** | **0.8735** | 0.9954 |

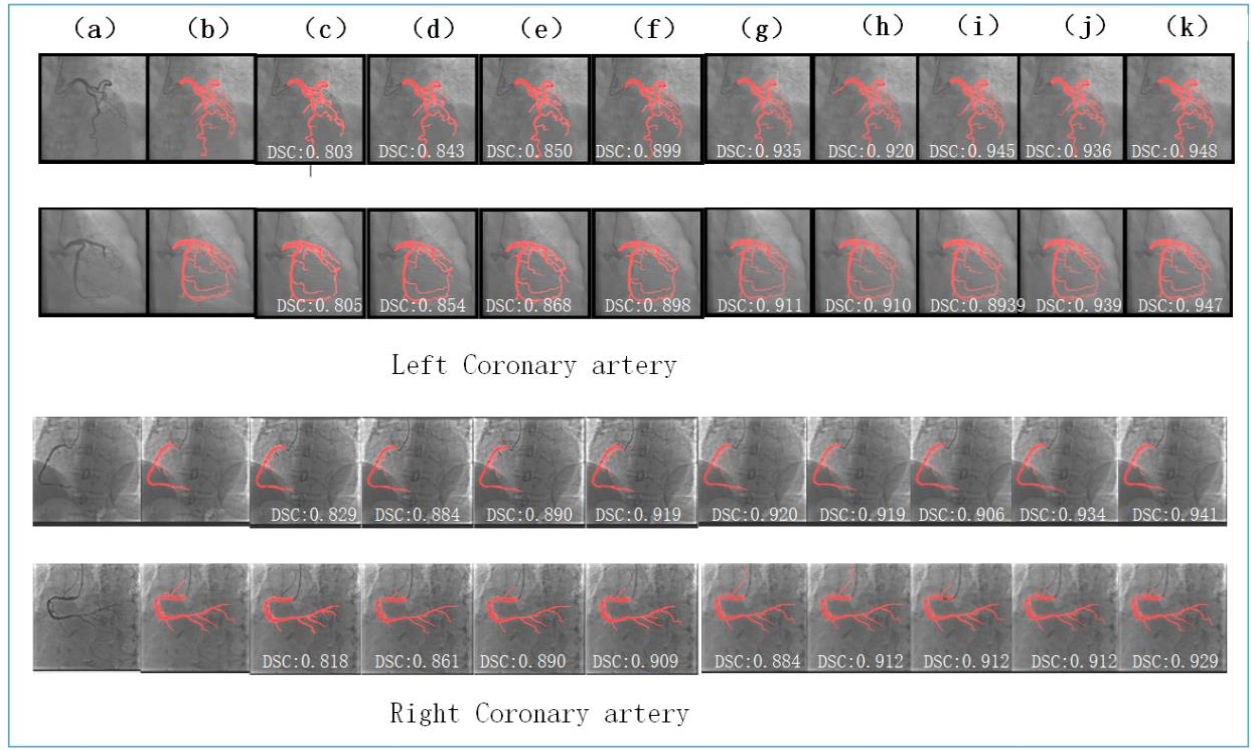

**Fig. 3** The raw ICAs and ground truth are shown in (a) and (b), respectively. Arterial segmentation results of the left coronary artery and the right coronary artery illustrated by (c) CGNet (d) PSPNet (e) DeepLabV3 (f) DeepLabV3+ (g) U-Net, (h) U-Net++ (i) U-Net3+w/o DS and TL, (j) U-Net3+w/o DS, (k) U-Net 3+ obtained for the arterial segmentation results.

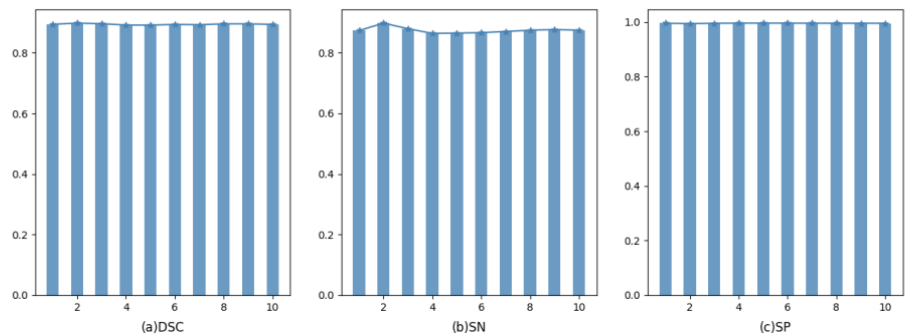

**Fig. 4**, Results of the 10-fold cross-validation experiment. (a) DSC (b) SN (C) SP.

## 4. Discussion

### 4.1 Performance analysis

Several excellent deep learning models have been developed and achieved remarkable results on image segmentation tasks.

U-Net has been widely used for the segmentation of medical images since it was proposed. The plain skip connections in U-Net require fusion of the same proportion of encoder and decoder feature maps. However, the same proportion of feature maps from the decoder and encoder networks are not semantically identical, which means he skip connections in U-NET are unnecessarily limited. Therefore, the performance of feature repair in the decoder path is affected, leading to impaired quality and efficiency of coronary segmentation. [30].

U-Net++ introduces nested skip connections. The primary function of this redesigned skip connection is further to enhance the utilization of low and deep-level features, which aim to reduce the missing semantic information of the feature maps in the encoding and decoding sub-networks [15]. U-net++ has excellent benefits for exploring multi-scale information with the nested dense skip connection [31]. However, short connections basically reprocess the decoding features, coupled with the fusion of various connections, and the original features of multi-scale information are hardly well utilized. In the research of segmentation tasks, high-level and low-level semantic feature maps can capture different information, respectively. However, some of the required information may be diluted in down and up-sampling[25].

PSPNet, which takes advantage of the pyramid parsing module and can aggregate contextual information from different regions. CGNet learn both local and global features by obtaining contextual texture features. DeepLabV3, which takes advantage of spatial pyramidal pooling based on atrous convolution to augment image-level features.

To improve the accuracy of image segmentation, the U-Net 3+, which utilizes full-scale skip connections and full-scale deep supervision, has been proposed [16]. The full-scale skip connection in U-Net 3+ can simultaneously combine the low-level semantic features and high-level abstract features in the down-sampling process, increasing the receptive field.

In this study, we use the U-Net 3+ [16] for accurate coronary artery extraction. For the coronary artery segmentation task, the full-scale skip connections enable exploring more semantic information and capturing more fine-grained details and improve our segmentation accuracy. Deep-supervised methods have shown an excellent capability in solving the vanishing gradient and exploding gradient problem inherent with CNNs [32],[33]. The full-scale deep supervision not only avoids over fitting, but also enables the CNN to learn a richer representation of arterial information, which is beneficial to the segmentation of arterial details. In addition, the U-Net 3+ can learn arterial semantic information from aggregated feature maps at different scales and enhance the extraction of arterial boundary information.

It can be concluded that U-Net3+, which we applied, obtained the highest DSC on the ICA dataset (Table 2). The average Dice score of U-Net 3+ was 1.43% and 1.28% higher than that of U-Net and U-Net++. The average Dice score of the U-Net 3+ model using transfer learning and full-scale deep supervision with hybrid loss functions increased from 0.8607 to 0.8942. The SN was 0.0430 higher than U-net and 0.0404 higher than U-Net++. In the evaluation of HD and ASD, Unet3+ obtained results of 6.0794 and 0.3247, respectively, which are better than U-Net ++ (HD of 6.1048 and ASD of 0.4588) and U-net (HD of 6.1040 and ASD of 0.4403).

As shown in Figure 5, the green indicates false negative segmentation, where the artery is incorrectly predicted as background. The red color indicates false positive segmentation where the background is incorrectly predicted as an artery. The red color indicates the artery that has been accurately extracted. The U-Net 3+ segmentation model proposed for in this paper predicts fewer erroneous pixel regions than other models.

The advantage of the tested network includes the following: it can achieve ideal DSC when applied to deal with ICAs and the feasibility of the tested method was demonstrated by ablation experiments.

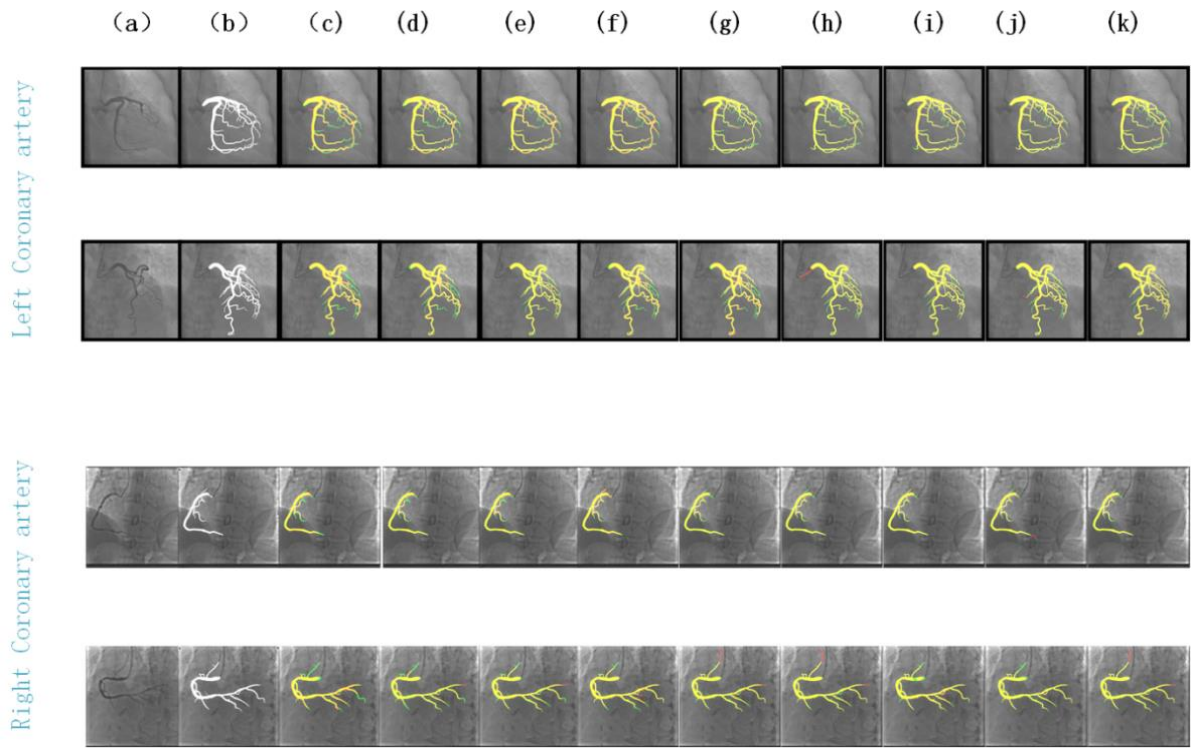

**Fig. 5** Comparison between the segmentation results and the ground truth. Green and red colors indicate the pixels with segmentation errors. (a) ICA, (b) ground truth, (c) CGNet (d)PSPNet I DeepLabV3 (f) DeepLabV3+ (g) U-Net, (h) U-Net++ (i) U-Net3+w/o DS and TL, (j) U-Net3+w/o DS, (k) U-Net 3+. The green, red and yellow regions represent false negative, false positive and true positive pixels, respectively.

### 4.2 Clinical overview and application

At present, ICA is still the reference gold standard imaging technique for the evaluation of clinically significant CAD. The fundamental task required for the interpretation of ICA is to identify arteries and quantify the severity of coronary stenosis. The general scheme for automated or semi-automated quantitative stenosis assessment in ICA is artery extraction, geometry calculation, and stenosis analysis [4]. Accurate coronary artery extraction is an important step in the stenosis detection task; however, this step is very challenging due to many complicated factors.

We have made continuous effort in automatic extraction and stenosis detection of coronary arteries from 2D ICAs [14][31]. Our recent clinical application includes reconstruction of 3D arterial anatomy from bi-planar 2D ICAs, which is then fused with myocardial functions measured from gated SPECT myocardial perfusion images for clinical decision making to improve the diagnosis and risk stratification of CAD [34][35]. Automatic and accurate extraction of coronary arteries play an essential role in these applications.

The proposed method in this study represents our recent advancement. It has shown improved performance, especially in patients with diffuse stenosis or continuous angiography, where accurate extraction is extremely important. It will further facilitate the assessment of coronary stenosis and thus has great promise for clinical use.

**4.3 Limitations**

There are the following limitations on automatic coronary artery extraction by applying U-Net3+.

1) For supervised deep learning, a large number of manually labeled samples are required to train a better deep learning model to obtain better results. More samples will further improve the performance of our method.

2) Automatic coronary artery extraction in this study is based on the contour of single-view single-frame ICA, so adjacent frames in the ICA videos will improve the performance of our method, particularly for missing and overlapped vessels. This will be incorporated in our future studies.

**Conclusion**

In this paper, we proposed a deep learning-based method for automatic extraction of coronary arteries from ICA. we trained the U-Net 3+ model for coronary artery extraction and validated it on our dataset. By employing transfer learning, the U-Net3+ model trained with the hybrid loss function could achieve excellent performance on the coronary artery extraction task. The method has great promise for clinical use.

## Acknowledgements

This research was supported by the Key Science Research Project of Colleges and Universities in Henan Province of China (No. 22A520046).